\documentclass{article}
\usepackage{spconfa4,amsmath,graphicx}

\usepackage[dvipsnames]{xcolor}
\usepackage[colorlinks=true,bookmarks=false]{hyperref}
\hypersetup{allcolors=Black}
\usepackage[capitalize,nameinlink]{cleveref}
\crefname{equation}{}{}
\Crefname{equation}{Equation}{Equations}

\usepackage{acronym}
\acrodef{ae}[AE]{autoencoder}
\acrodef{auc}[AUC]{area under the \ac{roc} curve}
\acrodef{asd}[ASD]{anomalous sound detection}
\acrodef{knn}[kNN]{$k$-nearest neighbor}
\acrodef{lora}[LoRA]{low-rank adaptation}
\acrodef{mlp}[MLP]{multilayer perceptron}
\acrodef{pauc}[pAUC]{partial \ac{auc}}
\acrodef{roc}[ROC]{receiver operating characteristic}
\acrodef{ssl}[SSL]{self-supervised learning}
\acrodef{sota}[SOTA]{state-of-the-art}
\acrodef{asr}[ASR]{automatic speech recognition}
\acrodef{tse}[TSE]{target sound extraction}
\acrodef{mse}[MSE]{mean squared error}
\acrodef{mhsa}[MHSA]{multi-head self-attention}
\acrodef{swiglu}[SwiGLU]{Swish gated linear unit}
\acrodef{ffn}[FFN]{feed-forward network}
\acrodef{sid}[SID]{speaker identification}
\acrodef{er}[ER]{emotion recognition}
\acrodef{map}[mAP]{mean average precision}
\acrodef{snr}[SNR]{signal-to-noise ratio}
\acrodef{tse}[TSE]{target sound extraction}
\acrodef{dbeats}[DBEATs]{denoising BEATs}
\acrodef{nabeats}[NABEATs]{noise-aware BEATs}
\acrodef{ss}[SS]{spectral subtraction}
\acrodef{na}[NA]{noise-aware}
\acrodef{ca}[CA]{cross-attention}
\usepackage{amssymb}
\usepackage{multirow}
\usepackage{makecell}
\usepackage{bm}
\usepackage{booktabs}
\usepackage{pifont}
\usepackage{color}
\usepackage{siunitx}
\usepackage{amsfonts}
\usepackage{adjustbox}
\usepackage{caption}
\usepackage{enumitem}

\usepackage[
  backend=biber,
  bibstyle=ieee,
  citestyle=numeric-comp,
  sorting=none,
  maxbibnames=10,
  doi=false,
  isbn=false,
  url=false,
  eprint=false
]{biblatex}
\defbibheading{bibliography}[\refname]{}

\AtBeginBibliography{%
  \footnotesize
  \setlength{\itemsep}{0.1\baselineskip}
  \setlength{\parskip}{0pt}
}
\AtEveryBibitem{%
  \ifentrytype{article}
    {}
    {\clearfield{pages}}%
}
\DeclareSourcemap{
	\maps[datatype=bibtex, overwrite=true]{
		\map{
		    \step[fieldsource=booktitle,
			match=\regexp{.*EUSIPCO.*},
			replace={Proc. EUSIPCO}]
			\step[fieldsource=booktitle,
			match=\regexp{.*CVPR.*},
			replace={Proc. CVPR}]
			\step[fieldsource=booktitle,
			match=\regexp{.*Interspeech.*},
			replace={Proc. Interspeech}]
			\step[fieldsource=booktitle,
			match=\regexp{.*ICASSP.*},
			replace={Proc. ICASSP}]
			\step[fieldsource=booktitle,
			match=\regexp{.*ICLR.*},
			replace={Proc. ICLR}]
			\step[fieldsource=booktitle,
			match=\regexp{.*IJCAI.*},
			replace={Proc. IJCAI}]
			\step[fieldsource=booktitle,
			match=\regexp{.*IJCNN.*},
			replace={Proc. IJCNN}]
			\step[fieldsource=booktitle,
			match=\regexp{.*ICML.*},
			replace={Proc. ICML}]
            \step[fieldsource=booktitle,
			match=\regexp{(?i).*international.*conference.*on.*machine.*learning.*},
			replace={Proc. ICML}]
			\step[fieldsource=booktitle,
			match=\regexp{.*ASRU.*},
			replace={Proc. ASRU}]
			\step[fieldsource=booktitle,
			match=\regexp{.*SLT.*},
			replace={Proc. SLT}]
			\step[fieldsource=booktitle,
			match=\regexp{.*SSW.*},
			replace={Proc. SSW}]
			\step[fieldsource=booktitle,
			match=\regexp{.*WASPAA.*},
			replace={Proc. WASPAA}]
			\step[fieldsource=booktitle,
			match=\regexp{.*DCASE.*},
			replace={Proc. DCASE}]
            \step[fieldsource=booktitle,
            match=\regexp{(?i).*Detection.*and.*Classification.*of.*Acoustic.*Scenes.*and.*Events.*Workshop.*},
			replace={Proc. DCASE}]
            \step[fieldsource=booktitle,
			match=\regexp{.*MLSP.*},
			replace={Proc. MLSP}]
            \step[fieldsource=booktitle,
			match=\regexp{.*ECCV.*},
			replace={Proc. ECCV}]
            \step[fieldsource=booktitle,
			match=\regexp{.*AAAI.*},
			replace={Proc. AAAI}]
            \step[fieldsource=booktitle,
			match=\regexp{.*ACM.*Multimedia.*},
			replace={Proc. ACM MM}]
            \step[fieldsource=booktitle,
			match=\regexp{.*NeurIPS.*},
			replace={Proc. NeurIPS}]
            \step[fieldsource=booktitle,
            match=\regexp{(?i).*advances.*in.*neural.*information.*processing.*systems},
            replace={Proc. NeurIPS}]
            \step[fieldsource=journal,
			match=\regexp{.*TASLP.*},
			replace={IEEE/ACM TASLP}]
			\step[fieldsource=journal,
			match=\regexp{(?i).*Transactions.*on.*Audio.*Speech.*and.*Language.*Processing.*},
			replace={IEEE/ACM TASLP}]
            \step[fieldsource=journal,
			match=\regexp{(?i).*Trans.*on.*Acoust.*Speech.*Sig.*Process.*},
			replace={IEEE TASSP}]
            \step[fieldsource=journal,
			match=\regexp{(?i).*Trans.*Affect.*Comput.*},
			replace={IEEE Trans. Affect. Comput.}]
            \step[fieldsource=journal,
			match=\regexp{(?i).*Open.*Journal.*Signal.*Processing.*},
			replace={IEEE OJSP}]
            \step[fieldsource=journal,
			match=\regexp{.*JAIR.*},
			replace={JAIR}]
            \step[fieldsource=journal,
			match=\regexp{.*J-STSP.*},
			replace={IEEE J-STSP}]                
			\step[fieldsource=journal,
			match=\regexp{(?i).*Selected.*Topics.*Signal.*Processing.*},
			replace={IEEE J-STSP}]                
			\step[fieldsource=journal,
			match=\regexp{(?i).*Journal.*Machine.*Learning.*Research.*},
			replace={JMLR}]                
			\step[fieldsource=booktitle,
			match=\regexp{(?i).*Proc.*Meetings.*Acoust.*},
			replace={POMA}]
			\step[fieldsource=series,
			match=\regexp{.+},
			replace={{}}]
			\step[fieldsource=editor,
			match=\regexp{.+},
			replace={{}}]
			\step[fieldsource=publisher,
			match=\regexp{.+},
			replace={{}}]
			\step[fieldsource=month,
			match=\regexp{.+},
			replace={{}}]
			\step[fieldsource=location,
			match=\regexp{.+},
			replace={{}}]
			\step[fieldsource=address,
			match=\regexp{.+},
			replace={{}}]
			\step[fieldsource=organization,
			match=\regexp{.+},
			replace={{}}]
		}
	}
}

\usepackage{tikz}
\usetikzlibrary{arrows.meta,positioning,fit,calc}
\usepackage{booktabs}
\usepackage{makecell}
\usepackage{tabularx}

\title{NABEATs: Noise-Aware Audio Representation Learning}
\name{{\shortstack[c]{Takuya Fujimura$^{1,2*}$\thanks{*This work was done during an internship at MERL.}, Yoshiki Masuyama$^{1}$, Gordon Wichern$^{1}$, \\ Christoph Boeddeker$^{1}$, Julius Richter$^{1}$, Jonathan Le Roux$^{1}$}}}
\address{$^1$Mitsubishi Electric Research Laboratories (MERL), Cambridge, USA,\\$^2$Nagoya University, Nagoya, Japan}
\begin{document}
\ninept
\maketitle
\begin{abstract}
We propose the concept of noise-aware audio \ac{ssl}, whose goal is to encode audio mixtures while suppressing undesired noise, and present Noise-Aware BEATs (NABEATs) as a BEATs-based realization of this framework.
Audio \ac{ssl} models are designed to handle a wide range of audio signals.
Consequently, under noisy conditions, they cannot effectively focus on the target sounds relevant to a downstream task, resulting in degraded performance.
To address this issue, NABEATs is trained to estimate clean BEATs representations from a noisy audio signal with an auxiliary reference noise input.
This reference noise enables the model to account for specific noise characteristics at inference time, thereby achieving better generalization across operating environments.
Our experimental evaluations demonstrate that NABEATs significantly improves performance of various downstream tasks under noisy conditions and also generalizes well to unseen noise types.
\end{abstract}
\acresetall


\section{Introduction}
\label{sec:intro}
\Ac{ssl} models have achieved remarkable success in speech and audio processing.
These models learn general-purpose representations from large amounts of unlabeled audio data, enabling transfer to various downstream tasks.
For example, speech \ac{ssl} models~\cite{hsu2021hubert,baevski2020wav2vec} have demonstrated impressive performance in downstream tasks such as automatic speech recognition, even with a small amount of labeled data or a simple model~\cite{yang21c_interspeech}.
Beyond speech, audio \ac{ssl} models~\cite{pmlr-v202-chen23ag,niizumi2022byol,niizumi2024masked,gong2022ssast} have also shown their effectiveness across various downstream tasks including the environmental sound and music domains.

Inspired by the success of \ac{tse}~\cite{delcroix2018single,liu2024separate,Kwon2025}, \ac{ssl} models that extract target representations from noisy signals have also been explored, especially in the speech domain~\cite{chen2022wavlm,fazel2023cocktail,wang2023adapter}.
A representative example is WavLM~\cite{chen2022wavlm}, which is trained to predict pseudo labels of the primary speaker's utterance and ignore both noise and non-primary speech.
In addition, speaker-aware variants~\cite{zhang23w_interspeech,huang2023adapting,lin2024sa} have also been proposed to extract the representation of a target speaker specified by a reference enrollment utterance.
These methods have made \ac{ssl} models more effective under real-world noisy conditions and have broadened their applicability.

Despite the progress in noise-robust \ac{ssl} models for speech, this direction has not yet been explored for general audio \ac{ssl} models.
In addition, audio \ac{ssl} models are expected to be applicable to a wider variety of downstream tasks than speech \ac{ssl} models. A method in the spirit of WavLM would thus not be practical, as some types of sounds may need to be preserved for some downstream tasks but suppressed in others.
Furthermore, unlike speaker-aware \ac{ssl}, relying on target sound information as reference is unrealistic for some downstream tasks: for example, in audio tagging, which aims to identify the sound class, providing the target sound is essentially equivalent to solving the task.
In contrast, even when the target sound is unknown, noise information can often be available, and using noise as a reference is more practical in real-world scenarios.

\begin{figure}[t]
    \centering
    \input{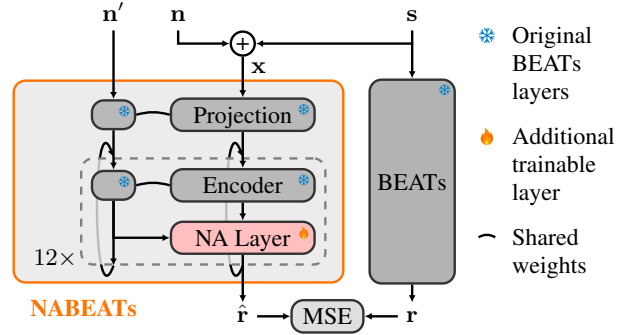}
    \caption{
        Overview of NABEATs. The model estimates clean BEATs representations $\mathbf r$ from noisy inputs $\mathbf x$ by leveraging an additional reference noise signal $\mathbf n'$ for conditional denoising.
    }
    \label{fig:overview}
\end{figure}

In this paper, we propose the concept of noise-aware audio \ac{ssl}, whose goal is to encode audio mixtures while suppressing undesired noise similar to a given reference.
As a specific realization, we introduce \ac{nabeats}, built upon the widely used and high-performing BEATs~\cite{pmlr-v202-chen23ag} model.
\ac{nabeats} is trained to estimate clean BEATs representations from noisy audio signals.
To achieve this, the model takes as additional input a reference noise signal $\mathbf n'$ sharing characteristics with the noise $\mathbf n$ in the noisy input (see \cref{fig:overview}) and learns to perform conditional denoising.
We also introduce \ac{dbeats} as a standard denoising baseline, which does not use the reference and is trained similarly to WavLM~\cite{chen2022wavlm}.
In the experimental evaluation, we demonstrate that both \ac{nabeats} and \ac{dbeats} significantly improve performance under noisy conditions across various downstream tasks, while \ac{nabeats} shows superior generalization to unseen noise.

\section{Proposed methods}
The intended usage scenario of \ac{nabeats} is similar to that of traditional spectral subtraction~\cite{boll1979suppression}, where a reference noise signal is obtained from noise-only segments, pre-recorded noise at the target location, or a separately placed microphone.
In this paper, we focus on the case where the reference noise is obtained from a different time segment from the same recording.
We consider this setting to be a practical and reasonable choice, and investigating other conditions is left to future work.
The proposed method is expected to handle non-stationary noise better than spectral subtraction, and avoids cascaded systems by operating directly in the representation space.

Our base model, BEATs, is an audio \ac{ssl} model that performs masked prediction in the discrete token space, where the tokens are obtained from mel-spectrogram patches using an acoustic tokenizer.
BEATs extracts a representation sequence $\mathbf r\!\in\!\mathbb{R}^{L\times 768}$ from mel-spectrogram patches through a projection layer and 12 Transformer encoder layers, where $L$ is the sequence length equal to the number of mel-spectrogram patches.
BEATs is trained to predict the correct tokens in the masked positions from $\mathbf r$, and the tokenizer is also iteratively updated through distillation from the BEATs model, leading to strong downstream performance.

\Cref{fig:overview} shows the training framework of \ac{nabeats}.
It is trained by distillation using the following \ac{mse} loss $\mathcal{L}_{\mathrm{MSE}}$ between the target representation $\mathbf  r$ from the frozen original BEATs and its estimate $\hat{\mathbf r}$ from \ac{nabeats}, given by
\begin{align}
\mathcal{L}_{\mathrm{MSE}}&=\frac{1}{L\cdot768} \left\| \mathbf{r} - \hat{\mathbf r} \right\|_F^2,\\
\mathbf r\!&=\!\mathrm{BEATs}(\mathbf s),\\
\hat{\mathbf r}\!&=\!\mathrm{NABEATs}(\mathbf x, \mathbf{n}^\prime),
\end{align}
where $\|\cdot \|_F$ denotes the Frobenius norm. Here, $\mathbf{s}$ and $\mathbf{x}=\mathbf{s}+\mathbf{n}$ denote target sound and noisy input sound, respectively, and
$\mathbf{n}^\prime$ is a reference noise as introduced in \cref{sec:intro}.
We also introduce \ac{dbeats} as a standard denoising baseline, which does not use the reference noise:
\begin{align}
\hat{\mathbf r}\!=\!\mathrm{DBEATs}(\mathbf x).
\end{align}
The training procedure of \ac{dbeats} is identical to that of \ac{nabeats}.

Following \cite{rouditchenko24_interspeech}, we construct \ac{dbeats} and \ac{nabeats} by inserting additional layers after every encoder layer of the original BEATs, and train only these layers while keeping the original BEATs components frozen. %
For \ac{dbeats}, we employ the following standard Transformer layer as the additional denoising layer:
\begin{align}
    \mathbf{z}_\mathbf{x} &\xleftarrow{} \mathbf{z}_\mathbf{x} + \mathrm{MHSA}(\mathrm{RMSNorm}(\mathbf{z}_\mathbf{x})), \label{eq:mhsa} \\
    \mathbf{z}_\mathbf{x} &\xleftarrow{} \mathbf{z}_\mathbf{x} + \mathrm{FFN}_\mathrm{SwiGLU}(\mathrm{RMSNorm}(\mathbf{z}_\mathbf{x})), 
    \label{eq:ffn}
\end{align}
where $\mathbf{z}_\mathbf{x}\!\in\!\mathbb{R}^{L\times 768}$ is the representation of the noisy input $\mathbf{x}$ from the previous layer, MHSA denotes multi-head self-attention~\cite{Vaswani2017attention}, RMSNorm denotes root mean square normalization~\cite{Zhang2019RMSNorm}, and $\mathrm{FFN}_\mathrm{SwiGLU}$ denotes a \ac{swiglu}-based \ac{ffn}~\cite{shazeer2020glu}.
For \ac{nabeats}, both $\mathbf{x}$ and $\mathbf{n}^\prime$ are passed through the same frozen BEATs components, and the additional \ac{na} layers refine the representation $\mathbf{z}_\mathbf{x}$ by leveraging the reference noise representation.
We consider two types of \ac{na} layers, based on \ac{ca} and FiLM~\cite{perez2018film}, and refer to the resulting models as \ac{nabeats}-CA and \ac{nabeats}-FiLM, respectively.
Both models replace only \cref{eq:mhsa} with \ac{ca} or FiLM while keeping the same \ac{ffn} as in \cref{eq:ffn}.
Specifically, \ac{nabeats}-CA employs the following \ac{ca} instead of \cref{eq:mhsa}:
\begin{align}
    \mathbf{z}_\mathbf{x} &\xleftarrow{} \mathbf{z}_\mathbf{x} + \mathrm{MHCA}\!\left(\mathrm{RMSNorm}\left(\mathbf{z}_\mathbf{x}\right), \mathrm{RMSNorm}\left(\mathbf{z}_{\mathbf{n}^\prime}\right)\right), \label{eq:ca}
\end{align}
where MHCA denotes multi-head \ac{ca}, with the first argument used as the query and the second argument used as the key and value, and $\mathbf{z}_{\mathbf{n}^\prime}$ is the representation of the reference noise $\mathbf{n}^\prime$.
\ac{nabeats}-FiLM applies the same feature-wise linear modulation to each element of the representation sequence, modifying the $l$-th element $\mathbf{z}_\mathbf{x}^l$  using the averaged reference noise representation $\bar{\mathbf{z}}_{\mathbf{n}^\prime}$:
\begin{align}
    \bar{\mathbf{z}}_{\mathbf{n}^\prime} &= \frac{1}{L}\sum_{l=1}^{L}\mathrm{RMSNorm}(\mathbf{z}^l_{\mathbf{n}^\prime}),\\
    \bm{\gamma} &= \mathrm{Affine}(\bar{\mathbf{z}}_{\mathbf{n}^\prime}) \in \mathbb{R}^{768}, \\
    \bm{\beta} &= \mathrm{Affine}(\bar{\mathbf{z}}_{\mathbf{n}^\prime}) \in \mathbb{R}^{768}, \\
    \mathbf{z}_\mathbf{x}^l &\xleftarrow{} \mathbf{z}_\mathbf{x}^l + \bm{\gamma} \odot \mathrm{RMSNorm}(\mathbf{z}_\mathbf{x}^l) + \bm{\beta},
\end{align}
where Affine denotes an affine layer, and $\odot$ denotes element-wise multiplication.
\ac{nabeats}-FiLM uses only global reference-noise information, whereas \ac{nabeats}-CA adaptively aggregates reference information across the sequence.

Another way to implement noise-aware \ac{ssl} is to use a cascaded architecture that first performs \ac{tse} and then feeds its output into the \ac{ssl} model. 
However, such a cascaded approach tends to be inefficient and incur higher computational cost~\cite{aihara2025sunac}; therefore, in this paper, we focus on an integrated \ac{ssl} model.

\section{Experimental setup}
We conduct two types of evaluations.
First, in \cref{sec:downstream}, we evaluate performance on various downstream tasks under simulated noisy conditions using a custom-designed setup.
For the reference noise $\mathbf{n}^\prime$ in \ac{nabeats}, we use a noise-only segment immediately preceding $\mathbf{n}$ from the same recording, with the same duration, for both \ac{nabeats} training and the downstream tasks.
Second, in \cref{sec:asd}, we evaluate on the DCASE 2025 Task 2~\cite{Nishida2025}, which addresses \ac{asd} for machine condition monitoring.
In this task, the dataset contains noisy machine sounds and separately recorded noise-only signals, aligning well with one of the intended applications of our \ac{nabeats} framework.

\begin{table}[t]
    \centering
    \caption{
    Datasets used for the training of \ac{dbeats} and \ac{nabeats}.
    Text in parentheses indicates the official split of each dataset used at each stage.
    }
    \vspace{-6pt}
    \begin{tabularx}{\columnwidth}{l>{\hsize=0.72\hsize}X >{\hsize=1.28\hsize}X}
        \toprule
        \bfseries Stage ~~~~~~~& \bfseries Target & \bfseries Noise \\
        \midrule
        Train & FSD50K (Train) & WHAM! (Train), DEMAND, QUT-NOISE \\
        Valid & FSD50K (Valid) & WHAM! (Valid) \\
        \bottomrule
    \end{tabularx}
    \label{tab:dataset_distil}
\end{table}

\begin{table}[t]
    \centering
    \caption{
    Noise datasets used for downstream evaluation.
    For WHAM! and MUSDB, we followed the official splits.
    The CHiME-3 row shows the recording IDs used for each split.
    }
    \vspace{-6pt}
    \begin{tabular}{llll}
        \toprule
        \bfseries Noise dataset & \bfseries Train & \bfseries Valid & \bfseries Test \\
        \midrule
        WHAM! (seen) & Train & Valid & Test\\
        CHiME-3 (unseen) & 040 \& 050 & 030 & 010 \& 020 \\
        MUSDB18 (unseen) ~~~~& Train & Valid & Test\\
        \bottomrule
        \end{tabular}
    \label{tab:noise_dataset_downstream}
\end{table}

\begin{figure*}[t]
    \centering
    \centerline{\includegraphics[width=\textwidth]{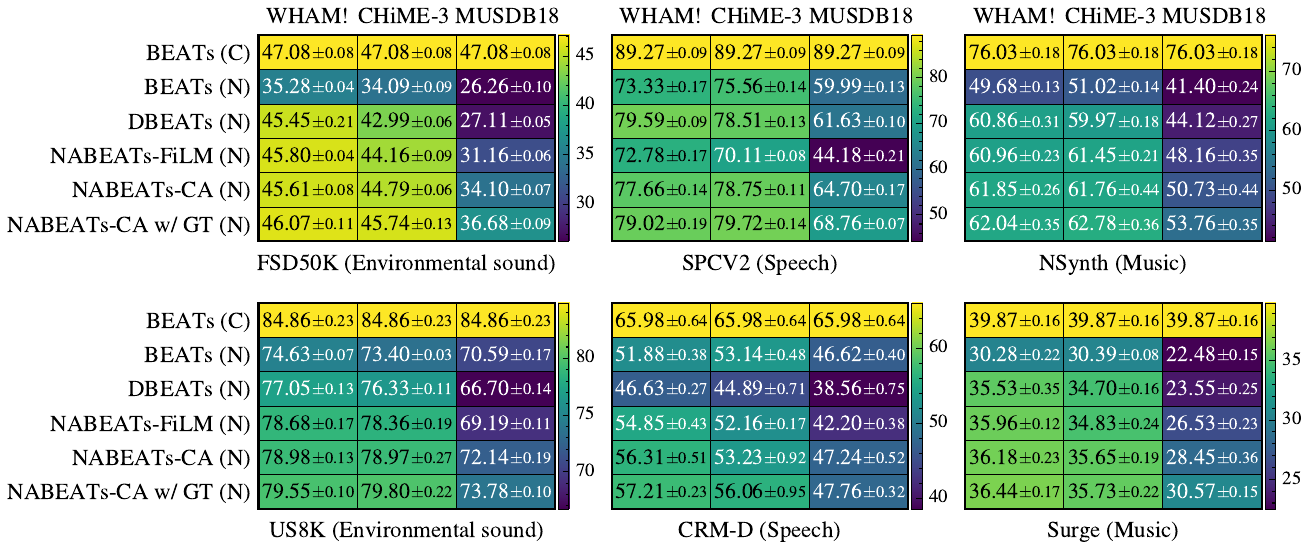}}
    \vspace{-6pt}
    \caption{Downstream classification performance under noisy conditions, where \acs{map} is used for FSD50K and accuracy is used for the other tasks.
    Each panel represents one task, with the columns corresponding to different noise conditions and the rows to different models.
    (C) and (N) denote clean and noisy input, respectively.
    Average scores across six trials with different random seeds are shown, together with 95\% confidence intervals.
    ``w/ GT'' indicates the performance when the reference noise for \ac{nabeats} is the ground-truth noise $\mathbf{n}$.
    }
    \label{fig:downstream}
\end{figure*}

\cref{tab:dataset_distil} summarizes the datasets for the training of \ac{nabeats} and \ac{dbeats}.
We used FSD50K~\cite{fonseca2022FSD50K} to sample the target sound.
Although BEATs itself is trained on the larger Audioset~\cite{audioset}, we leave such large-scale training to future work.
As the noise dataset, we used WHAM!48kHz~\cite{Wichern2019WHAM},  
following the official splits.
In addition, we used DEMAND~\cite{thiemann2013diverse} and QUT-NOISE~\cite{dean10_interspeech} as training noise datasets.
All of these noise datasets contain environmental sounds such as traffic, café, or park noise.
In our experiments, all audio signals were resampled to 16~kHz.
For both training and validation, we mixed a target sound with a noise signal at random SNRs between -5 and 10 dB.
During training, we randomly cropped or zero-padded the target sound to 10 sec.

For BEATs, we used the \textit{BEATs\_iter3.pt} checkpoint.\footnote{https://github.com/microsoft/unilm/tree/master/beats}
In the additional layers in \ac{dbeats} and \ac{nabeats}, we used MHSA and MHCA with 8 heads, respectively.
For \ac{ffn}, we set the hidden size to 2304.
We trained both \ac{dbeats} and \ac{nabeats} for 250 epochs using the AdamW optimizer, a fixed learning rate of 0.0001, and a batch size of 80.
We applied an exponential moving average with a decay rate of 0.999 after 20k training steps, and evaluated the model with the best validation loss.

\section{Evaluation on various downstream tasks}
\label{sec:downstream}
\subsection{Setups}
We conducted evaluation on various downstream tasks according to \cite{niizumi2022byol,niizumi2024masked}.
We trained only a single linear layer on top of the frozen \ac{ssl} model for downstream classification tasks.
We adopted six downstream tasks covering environmental sound, speech, and music domains.
For environmental sound classification, we used FSD50K and UrbanSound8K (US8K)~\cite{salamon2014dataset}.
For speech-related tasks, we used Speech Commands V2 (SPCV2)~\cite{warden2018speech} for word classification and CREMA-D (CRM-D)~\cite{cao2014crema} for emotion recognition.
For music tasks, we used NSynth~\cite{engel2017neural} for instrument family classification and the Surge Pitch Dataset (Surge)~\cite{turian2021one} for pitch classification.

For US8K, we conducted 10-fold cross-validation following the official splits.
During cross-validation, we randomly divided the non-test folds into training and validation sets with a ratio of 10:1.
For the other datasets, we followed the official training, validation, and test splits.
Following \cite{niizumi2022byol,niizumi2024masked}, for each dataset, audio signals were cropped or zero-padded to the average duration of the dataset. For metrics, we used \ac{map} for FSD50K, a multi-label task, and accuracy for all other single-label tasks~\cite{niizumi2022byol,niizumi2024masked}.

\cref{tab:noise_dataset_downstream} summarizes the noise datasets used to simulate noisy conditions for the above downstream tasks.
In addition to WHAM!, which was used to train \ac{nabeats} and \ac{dbeats}, we used CHiME-3~\cite{barker2015third} and MUSDB18~\cite{musdb18} to evaluate generalization to unseen noise types.
CHiME-3 contains environmental noise similar to the noise datasets used to train \ac{nabeats} and \ac{dbeats}, whereas MUSDB18 contains music signals that are substantially different from them.
We split CHiME-3 based on the recording IDs.
For MUSDB18, we used the official dataset parser\footnote{https://github.com/sigsep/sigsep-mus-db} to obtain the validation set.
For WHAM! and MUSDB18, we excluded signals shorter than 60 sec to ensure that reference noise can be obtained from the same recording across all downstream tasks, while CHiME-3 contains only long recordings.
For each split of the downstream datasets, we mixed each sample with a noise signal from the corresponding split of the noise dataset at random SNRs between -5 and 10 dB.

In the downstream tasks, we fed each representation averaged across the sequence into a linear layer~\cite{pmlr-v202-chen23ag}, which we trained for up to 200 epochs using the Adam optimizer, a learning rate of 0.0003, and a batch size of 200.
We used the validation set for early stopping with a patience of 20 epochs.
We conducted each evaluation six times with different random seeds, and report the average results with 95\% confidence intervals.

\begin{figure*}[t!]
    \centering
    \centerline{\includegraphics[width=\textwidth]{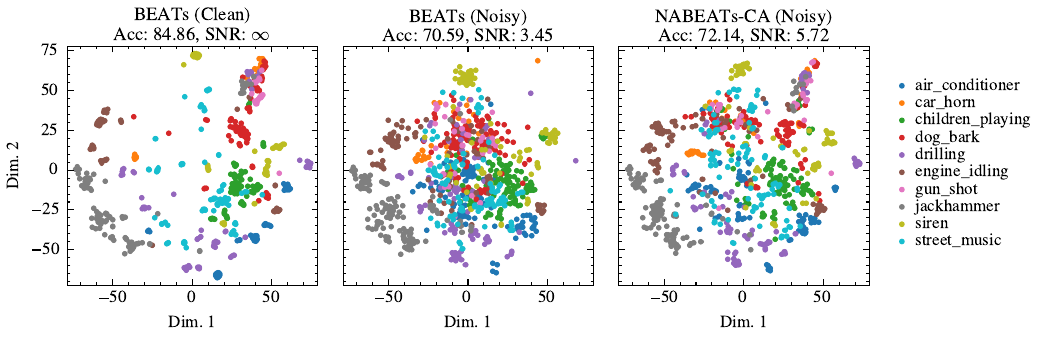}}
    \vspace{-6pt}
    \caption{t-SNE visualization of the averaged representation of the test set for US8K under the clean condition for BEATs and under the MUSDB18 noise condition for BEATs and NABEATs-CA.
    All figures have the same axes.
    }
    \vspace{-6pt}
    \label{fig:tsne}
\end{figure*}

\subsection{Results}
\Cref{fig:downstream} shows the results across various downstream tasks under noisy conditions, where we used noisy signals as input during both training and inference of the linear head.
First, we can see that performance of the original BEATs degrades significantly under noisy conditions compared with clean conditions in all tasks.
In contrast, \ac{nabeats}-CA and \ac{dbeats} improve the performance in the WHAM! and CHiME-3 noise conditions, except for \ac{dbeats} on CRM-D.
In the MUSDB18 noise condition, while \ac{dbeats} suffers from a large mismatch with the training noise data, \ac{nabeats}-CA still improves performance, demonstrating its better generalization to unseen noise conditions.
Furthermore, by leveraging reference noise, \ac{nabeats}-CA achieves better performance than \ac{dbeats} even in seen noise conditions in all tasks except SPCV2.
From the comparison between \ac{nabeats}-CA and \ac{nabeats}-FiLM, we can see that \ac{nabeats}-CA performs particularly well on the speech and music tasks and under the MUSDB18 noise condition, indicating that these tasks require more effective handling of time-varying information than environmental sound tasks.
Lastly, to investigate the behavior of \ac{nabeats}, we evaluated \ac{nabeats}-CA w/ GT, which uses the ground-truth noise $\mathbf{n}$ as the reference noise.
It yields a slight but consistent performance improvement over \ac{nabeats}-CA, which can be regarded as a reasonable upper-bound performance.
Since \ac{nabeats} is not trained using the ground-truth noise $\mathbf{n}$ as the reference noise, this result also shows the robustness of \ac{nabeats} to a different reference noise segment.

\Cref{tab:snr} shows the \ac{snr} between the clean representation sequences $\mathbf{r}$ and their estimates $\hat{\mathbf{r}}$, highlighting the substantial improvement by the proposed methods.
We note that \ac{dbeats} improves \ac{snr} even on CRM-D, despite degrading downstream performance, which indicates a potential mismatch between downstream performance and the \ac{mse} objective.
We leave investigation of better training objectives to future work.

\Cref{fig:tsne} shows t-SNE~\cite{van2008visualizing} visualizations of the averaged representation of the test set for US8K under the clean condition for BEATs and under the MUSDB18 noise condition for BEATs and NABEATs-CA.
The original BEATs representations are well separated across classes in the clean condition, but become mixed under the noisy condition.
In contrast, \ac{nabeats}-CA maintains better class separability under noisy conditions, leading to better downstream performance and \ac{snr}.

\begin{table}[t!]
\centering
\sisetup{
    reset-text-series = false, 
    text-series-to-math = true, 
    mode=text,
    tight-spacing=true,
    round-mode=places,
    round-precision=1,
    table-format=1.1,
    table-number-alignment=center
}
\caption{
SNR [dB] results between the clean and estimated representation sequences.
W and M indicate WHAM! and MUSDB18 noise conditions, respectively.
}
\vspace{-8pt}
\begin{tabular}{l*{6}{S}}
\toprule
& \multicolumn{2}{c}{FSD50K} & \multicolumn{2}{c}{CRM-D} & \multicolumn{2}{c}{NSynth} \\
\cmidrule(lr){2-3}\cmidrule(lr){4-5}\cmidrule(lr){6-7}
& W & M & W & M & W & M \\
\midrule
BEATs & 1.95 & 1.30 & 3.19 & 2.44 & 0.24 & -0.02 \\
DBEATs & 7.45 & 3.25 & 3.82 & 2.18 & 4.44 & 1.42 \\
NABEATs-CA & \bfseries 7.56 & \bfseries 5.39 & \bfseries 5.47 & \bfseries 4.05 & \bfseries 4.50 & \bfseries 3.19 \\
\bottomrule
\end{tabular}
\label{tab:snr}
\vspace{-3pt}
\end{table}

\vspace{-4pt}
\section{Evaluation on DCASE 2025 Task 2}
\vspace{-1pt}
\label{sec:asd}
We also conducted evaluation using the DCASE 2025 Task 2 dataset~\cite{Nishida2025}, compiled from~\cite{Harada2021,Dohi2022MIMIIDG,Harada2025}.
This dataset contains 15 machine types, each with 1000 normal training samples and 200 test samples including normal and anomalous samples.
Nine of the 15 machine types include 100 noise-only signals as supplementary data, and we restrict our experiments to these machine types.

Following \cite{saengthong2025deep}, without using any additional neural network, we simply calculated the anomaly score based on the distance between the representation of the test sample and those of the normal training samples, where we averaged the representation across the sequence.
We used the standard anomaly score calculation method which uses SMOTE oversampling~\cite{chawla2002smote} to mitigate the class imbalance in the training set and then calculates the cosine distance to the nearest training sample~\cite{Fujimura2025asdkit,jiang2025adaptive}.
For \ac{nabeats}, we selected a reference noise sample from the supplementary data with minimum distance to the input noisy machine sound in the original BEATs representation.

As the evaluation metric, we used the official task score based on the area under the receiver operating characteristic curve~\cite{Nishida2025} and report its harmonic mean across the nine machine types.

The evaluation scores on DCASE 2025 Task 2 are 54.57 for BEATs, 55.76 for \ac{dbeats}, and \textbf{56.15} for \ac{nabeats}-CA.
This further verifies the effectiveness of the proposed methods even when the reference noise is not the segment immediately preceding $\mathbf{n}$, and also demonstrates their effectiveness on a public benchmark dataset.

\vspace{-3pt}
\section{Conclusion}
\vspace{-3pt}
In this paper, we explored the concept of noise-aware audio \ac{ssl} and proposed \ac{nabeats} as a realization of this framework based on BEATs.
While the standard denoising baseline \ac{dbeats} is simply trained to estimate clean representations from noisy signals, \ac{nabeats} additionally takes a reference noise as input and is trained to suppress the corresponding noise information.
Our experiments showed that, while both methods significantly improve performance under noisy conditions, \ac{nabeats} outperforms \ac{dbeats} in both seen and unseen noise conditions, highlighting the benefit of using a reference noise to make the model noise-aware.
Future work includes investigating other reference noise settings, better training objectives for downstream tasks, and larger-scale training.

\clearpage
\section{References}
\label{sec:ref}
\printbibliography

\end{document}